\documentclass{PoS}

\usepackage[english]{babel} 

\usepackage{tabularx}
\usepackage[singlelinecheck=false]{caption}
\usepackage{booktabs}
\usepackage[caption=false]{subfig}
\usepackage[T1]{fontenc}
\usepackage{soul}
\usepackage{enumitem}
\usepackage{xcolor}
\usepackage[titletoc]{appendix}
\usepackage{overpic}
\usepackage[normalem]{ulem}
\usepackage[nointegrals]{wasysym}
\usepackage{tablefootnote}
\usepackage[caption=false]{subfig}
\usepackage{graphicx}
\usepackage{nicefrac}
\usepackage{tabularx}
\usepackage{booktabs}
\usepackage{multirow}
\usepackage{amssymb,amsmath}
\usepackage{footnote}
\usepackage{bm}
\usepackage[capitalize]{cleveref}[2012/02/15]

\usepackage{lineno}
\usepackage{accents}

\usepackage{lineno}

\usepackage{tabu}
\usepackage{url}
\usepackage{siunitx}
\DeclareSIUnit\parsec{pc}
\DeclareSIUnit\lightyear{ly}

\let\OldAng\ang%
\renewcommand*{\ang}[2][]{%
    \OldAng[scientific-notation=false,separate-uncertainty=true,round-mode=places,round-precision=2,#1]{#2}%
}
\DeclareSIUnit\year{yr}
\DeclareSIUnit\erg{erg}
\DeclareSIUnit\msun{M_{\astrosun}}
\DeclareSIUnit{\GeV}{\giga\electronvolt}
\DeclareSIUnit{\TeV}{\tera\electronvolt}
\DeclareSIUnit{\PeV}{\peta\electronvolt}
\DeclareSIUnit{\MeV}{\mega\electronvolt}
\DeclareSIUnit{\eV}{\electronvolt}
\DeclareSIUnit{\smm}{\square\metre\second}
\DeclareSIUnit{\smmr}{\metre^{-2}\second^{-1}}
\DeclareSIUnit{\dc}{d.c.}
\DeclareSIUnit{\pe}{p.e.}
\DeclareSIUnit{\nucleon}{nucleon}

\definecolor{desyOrange}{RGB}{242,142,0}
\usepackage{csquotes}
\usepackage[style=authoryear,giveninits=true,%
natbib=true,backend=bibtex,maxcitenames=2, style=numeric,
    sortcites,
    sorting=none,
doi=false,isbn=false,url=false,eprint=false]{biblatex}
\DeclareNameAlias{sortname}{last-first}

\usepackage{footnote}
\makesavenoteenv{tabular}
\makesavenoteenv{table}

\title{Modeling the non-thermal emission of the gamma Cygni Supernova Remnant up to the highest energies }

\ShortTitle{Modeling the non-thermal emission of the gamma Cygni Supernova Remnant}

\author{\speaker{Henrike Fleischhack}\\
        Michigan Technological University\\
        E-mail: \email{hfleisch@mtu.edu}}

\author{for the HAWC collaboration \thanks{Full author list and acknowledgements:  PoS(ICRC2019)1177 and \protect\url{http://www.hawc-observatory.org/collaboration/icrc2019.php}}}

\abstract{The gamma Cygni supernova remnant (SNR) is a middle-aged, Sedov-phase SNR in the Cygnus region. It is a known source of non-thermal emission at radio, X-ray, and gamma-ray energies. Very-high energy (VHE, >100 GeV) gamma-ray emission from gamma Cygni was first detected by the VERITAS observatory and it has since been observed by other experiments. Observations so far indicate that there must be a population of non-thermal particles present in the remnant which produces the observed emission. However, it is not clear what kind of particles (protons/ions or electrons) are accelerated in the remnant, how efficient the acceleration is, and up to which energy particles can be accelerated. Accurate measurements of the VHE gamma-ray spectrum are crucial to investigate particle acceleration above TeV energies. \\

This presentation will focus on multi-wavelength observations of the gamma Cygni SNR and their interpretation. We will present improved measurements of the VHE gamma-ray emission spectrum of gamma Cygni by the High-Altitude Water Cherenkov (HAWC) gamma-ray observatory, and use these results as well as measurements from other instruments to model the underlying particle populations producing this emission. HAWC's excellent sensitivity at TeV energies and above enables us to extend spectral measurements to higher energies and better constrain the maximum acceleration energy.  }

\FullConference{36th International Cosmic Ray Conference -ICRC2019-\\
		July 24th - August 1st, 2019\\
		Madison, WI, U.S.A.}

\begin{document}

\section{Introduction}
The $\gamma$ Cygni SNR (G78.2+2.1) is a middle-aged SNR located in the Cygnus region of our Galaxy, named for its location near the star $\gamma$ Cygni. The star is not related to the remnant; in the following text, ``$\gamma$ Cygni'' will be used to refer to the SNR. Its age has been estimated to be \SIrange{5000}{10000}{\year} \cite{2000AstL...26...77L,10.1093/mnras/stt1596}, and its distance from Earth to \SIrange{1.7}{2.6}{\kilo\parsec} \cite{10.1093/mnras/stt1596}.

Gamma-ray emission from the $\gamma$ Cygni SNR has been detected both in the GeV and the TeV range. The GeV emission has two components, an extended `disk' coincicent with the radio shell as well as a hotspot in the northwestern quadrant of the remnant (see e.g. 
\cite{Acero_2016, 3FGL, fges, Fraija:2016shx}). In the VHE regime, gamma-ray emission was first reported by VERITAS, which detected emission from the north-western hotpot, VER J2019+407 \cite{Aliu:2013yya}. The MAGIC collaboration has reported significant emission from both the hotspot and the disk component \cite{magicICRC}.

The Cygnus region, in which $\gamma$ Cygni is located, contains several gamma-ray sources. In particular, $\gamma$ Cygni overlaps with the Cygnus cocoon, an extended source of GeV to TeV gamma rays \cite{cocoon,cocoonFermi}. In the analysis of HAWC data, care must be taken to dis-entangle the two sources from each other, as not to over-estimate the emission from $\gamma$ Cygni. The multi-source analysis of the TeV emission from the region is described in \cite{cocoon}

In this study, we will extract the GeV to TeV energy spectrum of the $\gamma$ Cygni SNR and model it as a pion decay spectrum.

\begin{figure}[p]
\includegraphics[height=6cm]{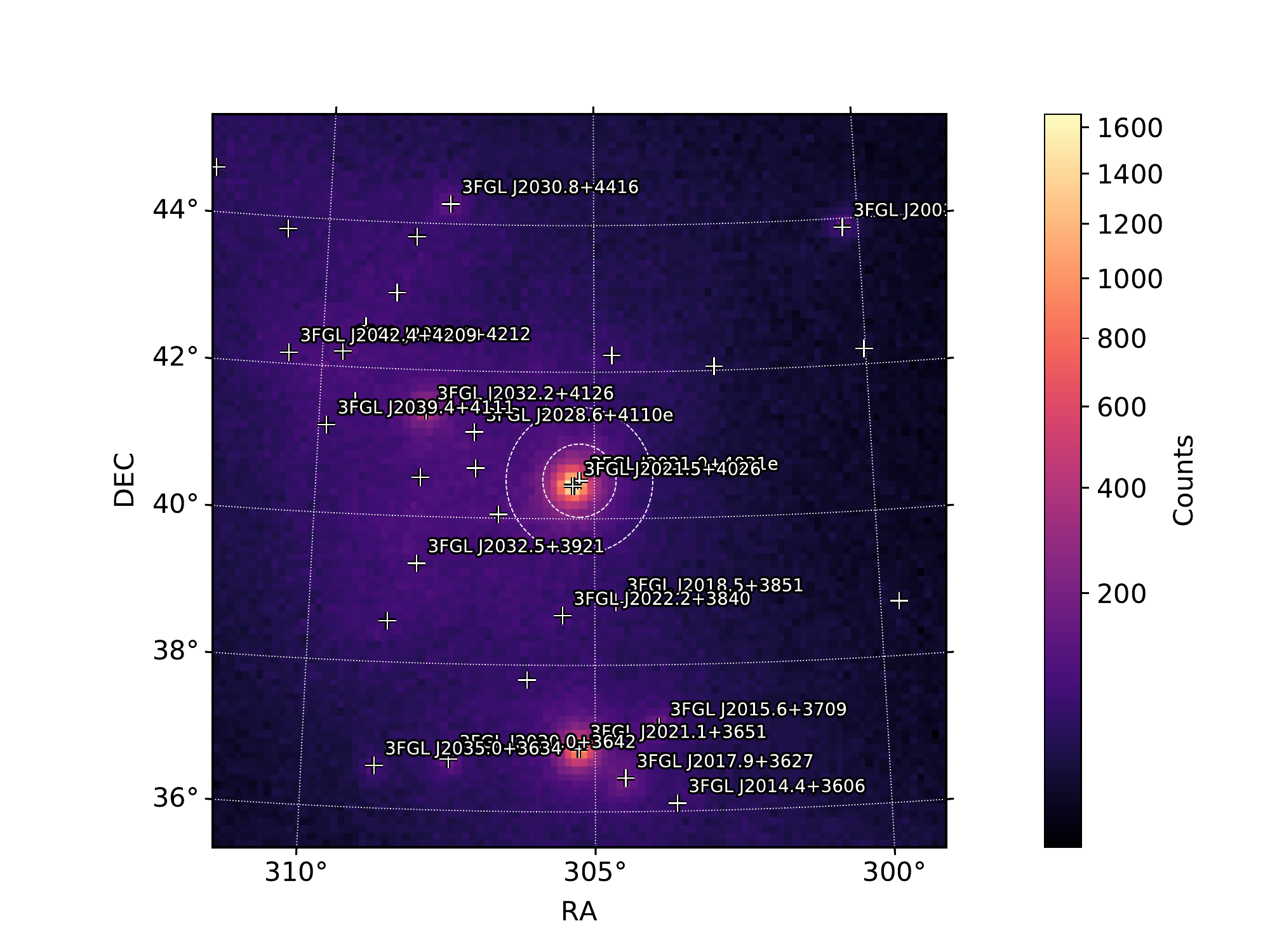}\includegraphics[height=6cm]{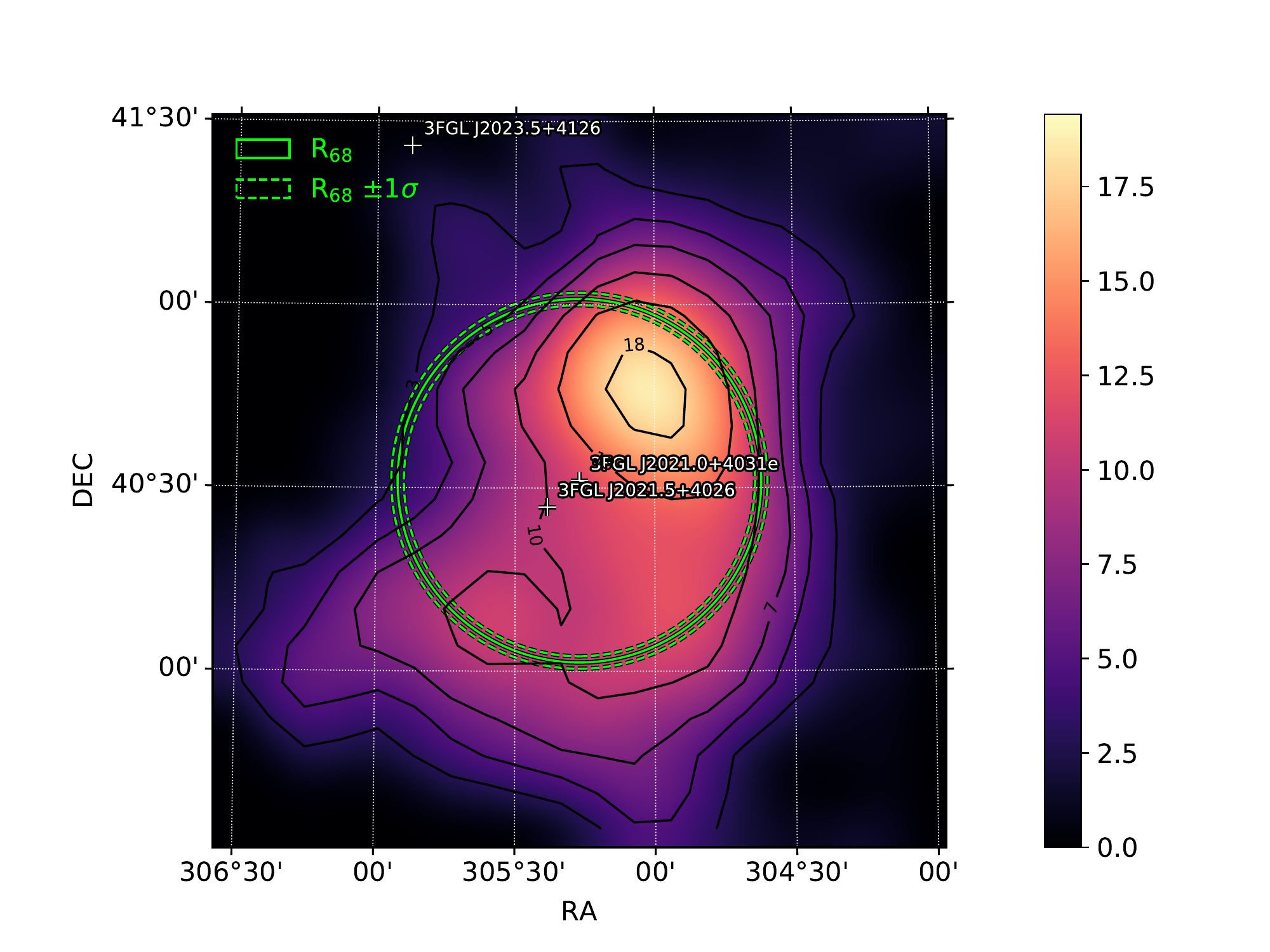}
\caption{\textit{Fermi}-LAT gamma-ray maps. Left: Gamma-ray counts above 1 GeV, smoothed. Right: TS map of $\gamma$ Cygni after subtracting all other sources in the region (point source assumption, spectral index -2). }
\label{fermimap}
\end{figure}

\begin{figure}[p]
\includegraphics[height=8cm]{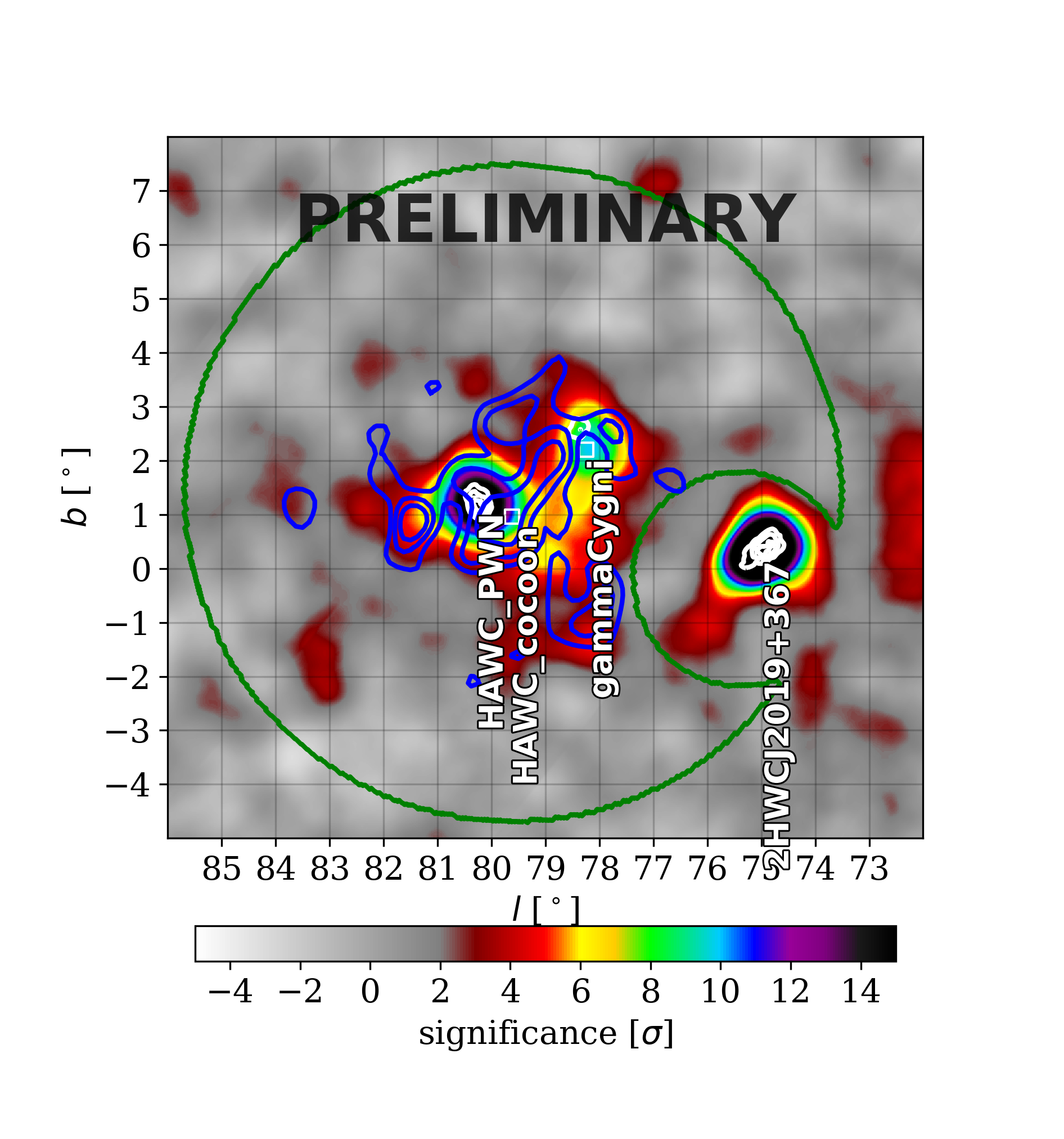}\includegraphics[height=8cm]{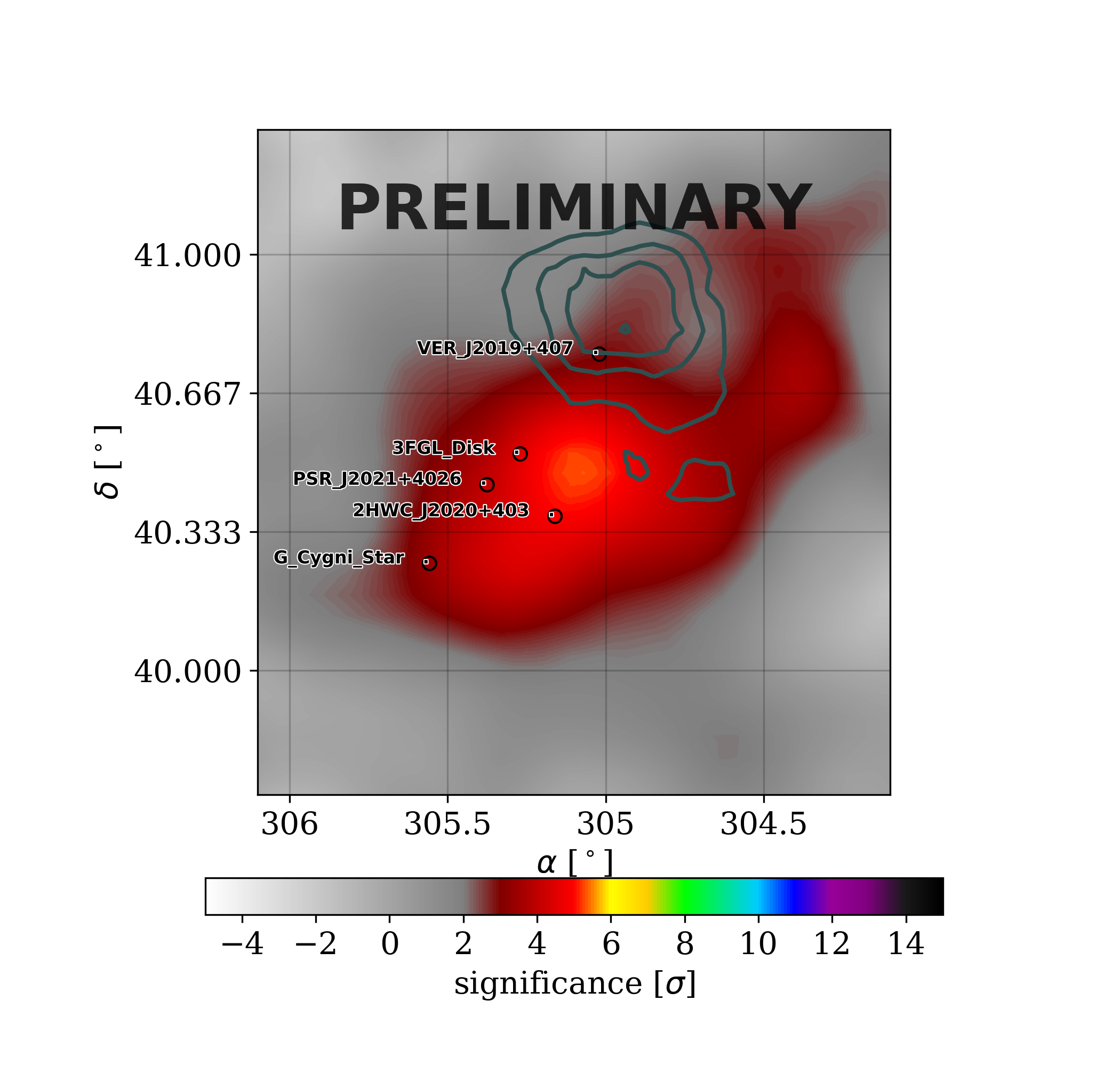}
\caption{HAWC gamma-ray maps. Left: significance map of the Cygnus cocoon region, with the ROI marked in green. Blue contours: Cygnus cocoon \cite{cocoonFermi}, White contours; VERITAS (\cite{Abeysekara_2018}, data provided by R. Bird). Right: Significance map of $\gamma$ Cygni after subtracting all other sources in the model (point source assumption, spectral index -2.7). Grey contours; VERITAS (\cite{Abeysekara_2018}.}
\label{hawcmap}
\end{figure}

\begin{table}[p]
\caption{Fit parameters of the energy spectra of the $\gamma$ Cygni SNR, fit to \textit{Fermi}-LAT and HAWC data separately. Statistical uncertainties only.}
\label{tab:spectra}
\begin{tabular*}{\textwidth}{@{\extracolsep{\fill} } llcccc}
\toprule
\textbf{Instrument} & \textbf{Component} & $E_0$ & $K$ & $\gamma$ & \textbf{TS} \\
 & & [TeV] &  [cm$^{-2}$s$^{-1}$TeV$^{-1}$] &  & \\
\midrule
\textit{Fermi}-LAT & Disk & 0.001 & \num[scientific-notation=true,separate-uncertainty=true]{1.2 \pm 0.1e-5} & \num[scientific-notation=false,separate-uncertainty=true]{2.06 \pm 0.03} & 580 \\
	& Hotspot &  0.001 &  \num[scientific-notation=true,separate-uncertainty=true]{0.28 \pm 0.06e-5} & \num[scientific-notation=false,separate-uncertainty=true]{2.11 \pm 0.09} & 142 \\
\midrule
\textit{Fermi}-LAT & Disk & 0.001 &  \num[scientific-notation=true,separate-uncertainty=true]{1.5 \pm 0.06e-5} & \num[scientific-notation=false,separate-uncertainty=true]{2.06 \pm 0.02} & 767 \\
\midrule
HAWC & Disk & 2.6 & \num[scientific-notation=true,separate-uncertainty=true]{3.0 \pm 0.5e-13} & \num[scientific-notation=false,separate-uncertainty=true]{3.02 \pm 0.14} & 39.3 \\
\bottomrule 
\end{tabular*}
\end{table}

\begin{figure}[p]
\includegraphics[width=\textwidth]{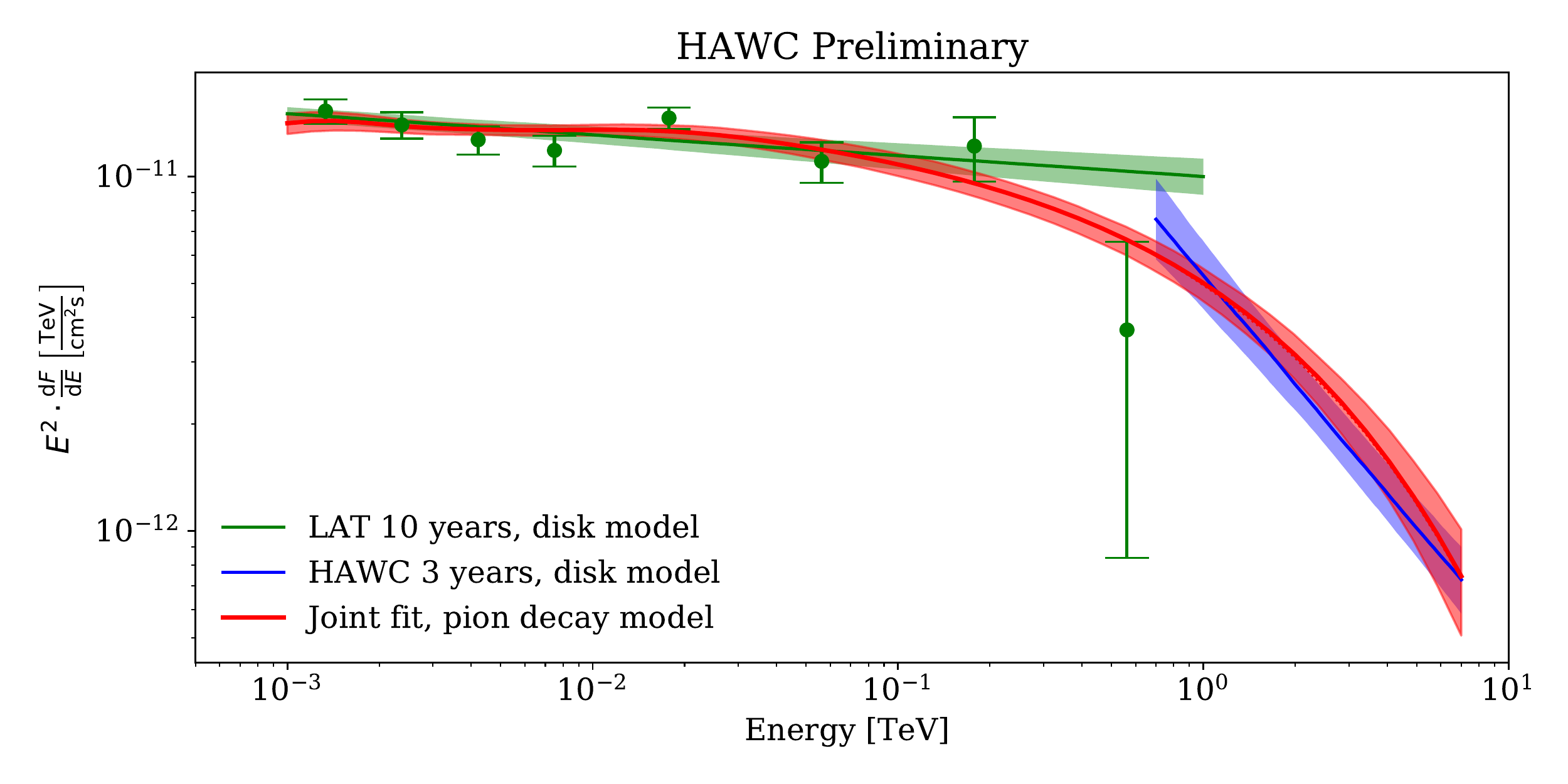}
\caption{Gamma-ray energy spectra of $\gamma$ Cygni. Statistical uncertainties only.}
\label{fig:spectra}
\end{figure}

\section{Instruments and Data Analysis}
Data from two instruments was used in this study: The \textbf{HAWC} (High Altitude Water Cherenkov) ground-based gamma-ray Observatory and the \textbf{\textit{Fermi}-LAT} (Large Area Telescope aboard NASA's \textit{Fermi} satellite).

HAWC is sensitive to gamma-ray air showers with a primary energy between a few hundred GeV to hundreds of TeV. Its angular resolution depends mainly on the fraction of the detector hit by a given event, and ranges from \ang{0.1} to \ang{1.0}. As an air shower detector, HAWC's main background is due to cosmic-ray induced showers which mimic gamma-ray events. HAWC's data are proprietary; details on data reconstruction and basic analysis can be found in \cite{2HWC, Abeysekara:2017mjj}. The dataset used for this analysis comprises 1038 days of HAWC data, reconstructed with the ground parameter energy estimator and binned according to the two-dimensional scheme described in \cite{Abeysekara:2019edl}. 

HAWC data are analyzed using a binned likelihood prescription as described in \cite{2015arXiv150807479Y}. For this study, the threeML framework \cite{Vianello:2015wwa} with the HAL plugin was used for HAWC data analysis as well as joint fits to HAWC and \textit{Fermi}-LAT data.

The LAT detects gamma rays from \SI{20}{\MeV} to several TeV \cite{2009ApJ...697.1071A,pass8,2013arXiv1303.3514A}. Its angular resolution ranges from several degrees below \SI{100}{\MeV} to about \ang{0.1} above \SI{30}{\GeV}\footnote{\url{https://www.slac.stanford.edu/exp/glast/groups/canda/lat_Performance.htm}}. The LAT has very good hadron rejection capabilities; for most analyses, the main backgrounds are gamma-ray photons from nearby sources or diffuse emission from the Galactic plane.

For this study, we used 9.8 years of \textit{Fermi}-LAT data (up to May 15, 2018), reconstructed according to pass 8 revision 2. \texttt{SOURCE} class events in between \SI{1}{\GeV} and \SI{870}{\GeV}, originating from the Cygnus region, were analyzed using the \texttt{fermipy}\cite{fermipy} analysis framework.

\section{Modeling of the Region}
For the analysis of the \textit{Fermi}-LAT data set, data within \ang{10} of the nominal position of the $\gamma$ Cygni SNR were selected. For the model, all 3FGL \cite{3FGL} sources within \ang{15} (accounting for photon leakage from sources outside the region of interest [ROI]) of the ROI center were considered, as well as the usual diffuse emission templates. All sources within \ang{5} of the ROI center and/or with at $TS>10$ in the initial \texttt{optimize} step were freed for the fit. Source positions and morphologies (for extended sources) were fixed to their 3FGL values. In particular, $\gamma$ Cygni (3FGL J2021.0+4031e) was modeled as a disk with centroid at RA \ang{305.27}, Dec \ang{40.52} and radius \ang{0.63}. 

The HAWC data mirrors what is described in \cite{cocoon}. The region of interest was chosen to be \ang{6} around RA=\ang{307.17} Dec=\ang{41.17}, with a \ang{2} cutout to remove contributions from the bright source 2HWC J2019+367. The HAWC emission is well described by three components. The extended TeV counterpart of the cocoon and the counterpart of VER J2031+415 are modeled as described in \cite{cocoon}, with their morphologies fixed and spectral parameters free. $\gamma$ Cygni was modeled with a disk-like morphology as in the 3FGL, with a power law energy spectrum. For this study, the morphology was fixed.

As in \cite{Fraija:2016shx}, we also considered the presence of a second `hotspot' component, a possible counterpart of VER J2019+407. For the \textit{Fermi}-LAT analysis, this was modeled with a gaussian morphology, centered on RA \ang{305.02}, Dec \ang{40.76}, with a 68\% containment radius of \ang{0.23}, and a power law energy spectrum. For the HAWC analysis, this source was modeled as a point source as its extend is small compared to HAWC's angular resolution, again with a power-law energy spectrum.

\section{Gamma-Ray Emission from the $\gamma$ Cygni SNR}

\subsection{Morphology}
In \textit{Fermi}-LAT data, both the `disk' and the `hotspot' component of the $\gamma$ Cygni SNR are significantly detected (TS of 142 for the hotspot and 581 for the disk). Their spectral indices are consistent with each other within uncertainties (see \cref{tab:spectra}). HAWC significantly detects the `disk' component, but there is no preference for including the hotspot. Significance maps derived from both datasets are show in \cref{fermimap} and \cref{hawcmap}.

\subsection{Energy Spectra}
In both the \textit{Fermi}-LAT and the HAWC data, $\gamma$ Cygni is well fit by power-law energy spectra, with no significant preference for curvature (see \cref{tab:spectra}, \cref{fig:spectra}). Spectral points were extracted for the \textit{Fermi}-LAT data range using \texttt{fermipy}'s \texttt{sed} function. $\gamma$ Cygni is detected with a TS of 39.3 in the HAWC dataset; this detection was not significant enough to extract spectral points.

The two spectra are plotted in \cref{fig:spectra}. They line up reasonably well with each other, but the TeV spectrum measured by HAWC is significanly softer than the GeV spectrum, indicating the presence of a spectral break or cutoff.

\section{Hadronic Modeling}
Assuming that the GeV--TeV emission from $\gamma$ Cygni is dominated by hadronic processes (relativistic protons and nuclei produce pions in interactions with the ISM, some of which decay into gamma rays), we attempted to extract the parameters of the underlying proton spectrum. The \texttt{naima} package \cite{naima} was used to predict the gamma-ray emission from a given proton spectrum. The \texttt{threeML} framework was used to fit the underlying model parameters to the \textit{Fermi}-LAT and HAWC data simultaneously. In this case, the HAWC data was read in directly via the \texttt{hawc\_hal} plugin, while the spectral points corresponding to the \textit{Fermi}-LAT measurement of $\gamma$ Cygni were read in via the \texttt{XYlike} plugin. The other two HAWC sources in the region were modeled as described above.

Pion decay emission from a simple power-law proton spectrum would produce gamma-ray emission following a pure power law above about 1 GeV, which is not compatible with the observed break in the gamma-ray spectrum. However, a powerlaw spectrum with an exponential cutoff, $\frac{dN}{dE} \propto \left(\frac{E}{E_0}\right)^{-\gamma} \cdot \exp\left(-\frac{E}{E_C}\right) $, reproduces the GeV to TeV data well. The best-fit proton model parameters is described by the following:
\begin{itemize}
\item Total energy $W_p$ in relativistic protons (above 1 GeV): \\ $W_p =  \left(1.4 \pm 0.1\right)\cdot 10^{50}\,\si{\erg} \cdot \left( \frac{\SI[scientific-notation=engineering]{2}{\parsec}}{d} \right)^2 \cdot \frac{\SI{1}{\per\cm\cubed}}{n}$, where $d$ is the distance to the SNR and $n$ is the number density of hydrogen nuclei in the emission region.
\item Proton spectral index $\gamma =  \num[scientific-notation=false,separate-uncertainty=true]{2.137 \pm 0.033}$
\item Proton cutoff energy: $E_C = \SI[scientific-notation=engineering,separate-uncertainty=true]{20\pm 5}{\TeV}$.
\end{itemize}

The uncertainties given here are statistical only. Not that there is no cross-calibration between \textit{Fermi}-LAT and HAWC yet; systematic differences in the flux or energy scales between the two instruments are under investigation and could lead to large uncertainties on the cutoff energy. The resulting gamma-ray energy spectrum can be seen in \cref{fig:spectra}.

These results are similar to what had been found in \cite{Fraija:2016shx}, which focussed on the hotspot component instead. 

\section{Conclusions and Discussion}
HAWC detects VHE gamma-ray emission from the $\gamma$ Cygni SNR, in spatial coincidence with the `disk' emission seen by \textit{Fermi}-LAT. The VHE energy spectrum is significantly harder than what is seen at GeV energies. The resulting GeV to TeV spectrum is well-fit by a pion decay model, assuming the underlying proton spectrum is well described by a power-law with an exponential cutoff.

In the future, we plan to repeat the \textit{Fermi}-LAT analysis of the region, using the new 4FGL catalog \cite{4FGL} and updated diffuse emission models as a baseline. In particular, we plan to look for a pion bump, a characteristic cutoff below 100 MeV indicative of emission from pion decays. 

Leptonic emission models are also under consideration.

\clearpage

\setlength\bibitemsep{0.2\baselineskip}
\printbibliography

\end{document}